\documentclass[12pt]{article}
\usepackage{putex}
\usepackage{graphicx}

\begin{document}

\preprint{PUPT-2322}

\institution{PU}{Joseph Henry Laboratories, Princeton University, Princeton, NJ 08544, USA}

\title{Notes on the bulk viscosity of holographic gauge theory plasmas}

\authors{
Amos Yarom\footnote{e-mail: {\tt ayarom@Princeton.EDU}}
}

\abstract{A novel technique is used to compute the bulk viscosity of high temperature holographic gauge theory plasmas with softly broken conformal symmetry. Working in a black hole background which corresponds to a non-trivial solution to the Navier-Stokes equation, and using a Ward identity for the trace of the stress-energy tensor, it is possible to obtain an analytic expression for the bulk viscosity. This can be used to verify the high temperature limit of a conjectured bound on the bulk viscosity for these theories. The bound is saturated when the conformal symmetry-breaking operator becomes marginal.}


\maketitle


\section{Introduction and summary}

The gauge-gravity duality \cite{Maldacena:1997re, Gubser:1998bc,Witten:1998qj} has provided an invaluable tool for studying the strongly coupled limit of large $N$ gauge theories. Perhaps surprisingly, there are features of these gauge theories which may be relevant to the quark-gluon-plasma (QGP) created at the Relativistic Heavy Ion Collider (RHIC) at the Brookhaven National Laboratory. In particular, it seems that the shear viscosity to entropy density ratio of the QGP formed at RHIC is of the order of \cite{Romatschke:2007mq}, 
\begin{equation}
\label{E:etas}
	\frac{\eta}{s} = \frac{1}{4\pi}
\end{equation}
with $\eta$ the shear viscosity and $s$ the entropy density.
The result in \eqref{E:etas} was first computed in \cite{Policastro:2001yc} and was later shown to hold in a large class of gauge theories with a holographic dual \cite{Buchel:2003tz,Kovtun:2004de,Buchel:2004qq}.  Similar relations exist for the electrical conductivity \cite{Kovtun:2008kx} and for higher order transport coefficients \cite{Haack:2008xx}. In a conformal gauge-theory plasma the bulk viscosity, $\zeta$, vanishes, but in a non-conformal theory the bulk viscosity is, in general, non-zero. In \cite{Buchel:2007mf} it was conjectured that the bulk viscosity always satisfies
\begin{equation}
\label{E:zeta}
	\frac{\zeta/\eta}{\frac{1}{d-1}-c_s^2} \geq 2
\end{equation}
with $d$ the number of space-time dimensions and $c_s$ the speed of sound,
for large $N$, strongly coupled gauge theories with a holographic dual. The bound \eqref{E:zeta} was shown to be satisfied in \cite{Mas:2007ng,Buchel:2007mf,Buchel:2008uu,Springer:2008js,Kanitscheider:2009as,David:2009np} for various configurations. In \cite{Gubser:2008sz} an explicit setup where the bound is slightly violated was constructed. It is not clear if the bulk matter content of this particular example follows from a truncation of string theory.

In this note, I show that in gauge theories with holographic duals involving gravity and a scalar which softly breaks the conformal symmetry, the bulk viscosity $\zeta$ and speed of sound $c_s$ are given by
\begin{align}
\label{E:csresult}
	 \frac{1}{d-1}  - c_s^2 & =   (d-2\Delta)\tan\left(\frac{\pi\Delta}{d}\right) D(\Delta,d) \left(\frac{\Lambda}{T}\right)^{2(d-\Delta)} \left(1+\mathcal{O}\left(\frac{\Lambda}{T}\right) \right) \\ 
\label{E:zetaresult}
	\left(\frac{d}{4\pi T}\right)^{d-1}\zeta  & = 2 \pi (d-\Delta)\, D(\Delta,d)\left(\frac{\Lambda}{T}\right)^{2(d-\Delta)}  \left(1+\mathcal{O}\left(\frac{\Lambda}{T}\right) \right)\,,
\end{align}
where
\begin{equation}
	D(\Delta,d) = \frac{16^{1-\Delta/d} (d-\Delta) d^{2(d-\Delta-1)}}{2(4\pi)^{2(d-\Delta)} (d-1)^2 } \left(\frac{\Gamma\left(\frac{\Delta}{d}\right)}{\Gamma\left(\frac{\Delta}{d}+\frac{1}{2}\right)}\right)^2\,,
\end{equation}
$\Delta$ is the scaling dimension of an operator dual to a bulk scalar, and $\Lambda^{d-\Delta}$ is the source term for this operator which is what breaks conformal invariance. It is assumed that $(d-2)/2 \leq \Delta < d$ so the scalar operator is relevant and above the unitarity bound. An expression similar to \eqref{E:csresult} was obtained in \cite{Cherman:2009tw,Hohler:2009tv} for $d=4$ and $2 < \Delta<4$ and expression \eqref{E:zetaresult} restricted to  $d=4$ and $2 \leq \Delta<4$ was conjectured in \cite{Cherman:2009kf}, based on a numerical analysis.\footnote{In \cite{Cherman:2009tw,Hohler:2009tv,Cherman:2009kf} the normalization of the scalar operator which breaks conformal invariance was such that its two point function vanishes in the limit $\Delta \to d/2$. This also results in a vanishing speed of sound in this limit. The conventions in this work follow that of \cite{Klebanov:1999tb}. To go from the conventions of  \cite{Cherman:2009tw,Hohler:2009tv,Cherman:2009kf} to the ones in \cite{Klebanov:1999tb} one should divide the source and expectation value of the scalar operator by $\Delta-d/2$. See section 2.2 of \cite{Klebanov:1999tb} for details.} 
Since \eqref{E:zetaresult} is valid in the high temperature limit, in theories with multiple scalars one should add a contribution of the form \eqref{E:zetaresult} for each scalar. The expressions for the bulk viscosity and speed of sound agree with the numerical results in the literature \cite{Benincasa:2005iv}.\footnote{In particular, one should make the identification $\Lambda^2 = 2 m_f^2$ or $\Lambda^4 = \frac{1}{6}m_b^4$ for the fermionic and bosonic deformations discussed in \cite{Benincasa:2005iv}.}

Using \eqref{E:csresult} and \eqref{E:zetaresult} and computing the leading contribution to the shear viscosity $\eta$, 
\[
	\eta = \left( \frac{d}{4\pi T}\right)^{d-1}\left(1 + \mathcal{O}\left(\frac{\Lambda}{T}\right)\right)\,,
\]
one finds that
\begin{equation}
	\lim_{T \to \infty} \frac{\zeta/\eta}{\left(\frac{1}{d-1}-c_s^2\right)} =\frac{d-\Delta}{d-2\Delta} 2\pi \cot\left(\frac{\pi \Delta}{d} \right) \geq 2.
\end{equation}
Curiously, the bound \eqref{E:zeta} is saturated in the limit where the conformal dimension of the scalar operator becomes marginal, $\Delta= d$. 

\section{Computation of the bulk viscosity}
In the hydrodynamic approximation the dynamics of a  fluid are specified by a velocity field $u^{\mu}(x)$ (canonically normalized so that $u^{\mu}u_{\mu}=-1$), an energy density $\epsilon(x)$ and  chemical potentials $\mu(x)$. In what follows the chemical potentials will be set to zero.  The energy momentum tensor of viscous hydrodynamics takes the form
\begin{equation}
\label{E:Tmnlinear}
	T^{\mu\nu} = \epsilon u^{\mu}u^{\nu} + P P^{\mu\nu} - \eta \sigma^{\mu\nu} - \zeta P^{\mu\nu}\partial_{\alpha}u^{\alpha} + \mathcal{O}(\partial^2)
\end{equation}
where $P$ is the pressure, related to the energy density through the equation of state, $\eta$ is the shear viscosity and $\zeta$ is the bulk viscosity. 
The projection operator $P^{\mu\nu}$ is given by
\begin{equation}
	P^{\mu\nu} = \eta^{\mu\nu} + u^{\mu}u^{\nu}
\end{equation}
and $\sigma^{\mu\nu}$ is given by
\begin{equation}
	\sigma^{\mu\nu} = P^{\mu\alpha}P^{\nu\beta} \left(\partial_{\alpha} u_{\beta} + \partial_{\beta} u_{\alpha} \right) - \frac{2}{d-1} P^{\mu\nu} P^{\alpha\beta}\partial_{\alpha}u_{\beta}\,.
\end{equation}
Greek indices run from 0 to $d-1$ and are raised and lowered with the Minkowski metric $\eta = \hbox{diag}(-1,1,\ldots,1)$.
By $\mathcal{O}(\partial^2)$ I mean terms which contain two derivatives of the hydrodynamic variables, for example $\sigma_{\mu\alpha}\sigma^{\alpha}_{\phantom{\alpha}\nu}$. See \cite{Baier:2007ix} for a detailed explanation of higher derivative terms and the derivation of \eqref{E:Tmnlinear}. 

In a conformally invariant theory the stress-energy tensor is traceless,
$
	T^{\mu}_{\mu} = 0\,,
$
which implies $\epsilon = (d-1) P$ and $\zeta = 0$. In this note, conformal invariance will be broken by sourcing a relevant operator $O_{\Delta}$ of conformal dimension $\Delta$, i.e, by adding to the boundary theory action a term $\frac{1}{\Delta-\frac{d}{2}}\int \Lambda^{d-\Delta} O_{\Delta} d^{d}x$.\footnote{As in \cite{Klebanov:1999tb}, the factor $(\Delta-d/2)^{-1}$ has been introduced in order to have a non vanishing vacuum expectation value for $O_{\Delta}O_{\Delta}$ in the $\Delta \to d/2$ limit.} Once the theory is not conformal,  the trace of the stress-energy tensor is given by
\begin{equation}
\label{E:Trace}
	T^{\mu}_{\mu} = -\frac{\Delta-d}{\Delta-\frac{d}{2}} \langle O_{\Delta} \rangle \Lambda^{d-\Delta} + \mathcal{A}\,.
\end{equation}
The relation \eqref{E:Trace} can be derived by holographic means, as in \cite{Bianchi:2001kw}, or by computing the beta-function for the source term for $O_{\Delta}$, as in \cite{Petkou:1999fv}. For certain $\Delta$ there will be an anomalous contribution, $\mathcal{A}$, to the trace of the stress tensor. Since $\mathcal{A}$ is independent of $\mathcal{O}$ it will not contribute to the speed of sound or bulk viscosity and will be ignored in what follows.
Comparing \eqref{E:Trace} with \eqref{E:Tmnlinear} one finds that
\begin{equation}
\label{E:zetaidentity}
	-\epsilon + (d-1) P - (d-1) \zeta \partial_{\alpha}u^{\alpha} + \mathcal{O}(\partial^2)= -\frac{\Delta-d}{\Delta-\frac{d}{2}} \langle O_{\Delta} \rangle \Lambda^{d-\Delta}\,,
\end{equation}
so the bulk viscosity can be computed by extracting the term proportional to the gradient of the velocity field from the right hand side of \eqref{E:zetaidentity}. Such a computation will be carried out in the remainder of this work for a class of strongly coupled gauge-theory plasmas in the planar limit, which can be described using the gauge-gravity duality. 

In detail, consider theories whose dual bulk action is given by
\begin{equation}
\label{E:action}
	S = \frac{1}{2\kappa^2}\int \sqrt{-g} \left( R - (d-1)d - \frac{1}{2}\partial\phi \partial\phi + V(\phi) \right) d^{d+1}x
\end{equation}
and where conformal invariance is softly broken by sourcing the operator $O_{\Delta}$ dual to the field $\phi$. Working in a coordinate system where the line element near the asymptotically AdS boundary (located at $r \to \infty$) takes the form
\begin{equation}
	ds^2 = \frac{1}{r^2} \left( -dv^2+ \sum_{i=1}^{d-1} (dx^i)^2\right) + 2 dv dr\,,
\end{equation}
the near boundary series expansion of the bulk field $\phi$ is given by \cite{Klebanov:1999tb}
\begin{equation}
\label{E:expansion}
	\phi = \frac{\langle O_{\Delta} \rangle}{(2 \Delta - d)} r^{-\Delta}\left(1+\mathcal{O}(r^{-1})\right) +\frac{\Lambda^{d-\Delta}}{\Delta-\frac{d}{2}}  r^{\Delta-d}\left(1+\mathcal{O}(r^{-1})\right)+ \ldots
\end{equation}
where $m^2 = 2 V''(0) = \Delta(d-\Delta)$. The expectation value of the operator $O_{\Delta}$ dual to $\phi$ is given by $\langle O_{\Delta} \rangle$, while $\Lambda^{d-\Delta}$ is a source term for $O_{\Delta}$.  Once a solution to the equations of motion following from \eqref{E:action} is found, the expectation value of $O_{\Delta}$ can be read off of \eqref{E:expansion}.

A thermal state of the boundary theory corresponds to a black hole geometry \cite{Witten:1998zw}. In the large temperature limit, one can keep only the leading terms in a $\Lambda/T$ expansion of the solution to the equations of motion. The near boundary behavior of the scalar field $\phi$ depends linearly on $\Lambda^{d-\Delta}$ so to leading order one is allowed to keep only terms linear in $\phi$. Thus, the Einstein equations are simply
\begin{equation}
\label{E:Einstein}
	R_{mn} = -d g_{mn}
\end{equation}
where $m,\,n = 0,\ldots,d$,
and the equation of motion for the scalar is
\begin{equation}
\label{E:scalareom}
	 \nabla^2\phi -m^2 \phi= 0.
\end{equation}
So the problem reduces to finding a black hole solution to the Einstein equations in the presence of a negative cosmological constant and finding a scalar field propagating in such a black hole background. It should be noted that this analysis can be easily generalized to several scalar fields: if the kinetic terms for the scalars are canonical then keeping only quadratic terms in the scalar potential amounts to decoupling the scalars from one another.

To obtain the high temperature limit of the bulk viscosity of these theories I use \eqref{E:zetaidentity} and compute the expectation value of $O_{\Delta}$ in a black hole background which is dual to a configuration with $\partial_{\alpha}u^{\alpha} \neq 0$. But before tackling this problem, it will be instructive to compute a simpler quantity,  the speed of sound, $c_s$. In a stationary configuration, where the spatial components of the velocity field vanish, one can use $c_s^2 = dP/d\epsilon$, to obtain
\begin{equation}
\label{E:tracesound}
	\left(\frac{1}{d-1}-c_s^2\right) = -\frac{\Delta-d}{(d-1)\left(\Delta-\frac{d}{2}\right)} \Lambda^{d-\Delta} \partial_{\epsilon} \langle O_{\Delta} \rangle
\end{equation}
from \eqref{E:zetaidentity}.
In order to evaluate the right hand side of \eqref{E:tracesound}, one needs to compute the expectation value of the operator $\langle O_{\Delta} \rangle$ dual to $\phi$ in a background dual to a stationary thermal state. Such a background is given by the AdS-Schwarzschild black hole solution whose line element is
\begin{equation}
\label{E:AdSSS}
	ds^2 = \frac{L^2}{r^2} \left( -f(r) dv^2+ \sum_{i=1}^{d-1} (dx^i)^2\right) + 2 dv dr
\end{equation}
where 
\begin{equation}
	f(r)=1-\frac{1}{(b r)^{d}}
\end{equation} 
and $b= d/(4 \pi T)$ with $T$ the Hawking temperature of the black hole, and also the temperature of the thermal state in the boundary theory. The equation of motion for the scalar in this background is
\begin{equation}
\label{E:scalareomO0}
	\frac{ \partial_r \left( r (1-(r b)^{d}) \partial_r \phi \right)}{r^{d-1} b^{d-2}} +\Delta(\Delta-d)\phi = 0\,.
\end{equation}
The solutions to \eqref{E:scalareomO0} are
\begin{equation}
	\phi = \mathbf{P}_{-1+\frac{\Delta}{d}} \left(-1+2 (r b)^{d} \right)
	\quad \hbox{and} \quad
	\phi = \mathbf{Q}_{-1+\frac{\Delta}{d}} \left(-1+2 (r b)^{d} \right)
\end{equation}
where $\mathbf{P}$ and $\mathbf{Q}$ are Legendre functions of the first and second kind. The solution to \eqref{E:scalareomO0} which is finite everywhere except at the black hole singularity and takes the asymptotic form \eqref{E:expansion} near the boundary is
\begin{equation}
\label{E:solO0}
	\phi(r b) = (\Lambda b)^{d-\Delta} \frac{2\Gamma\left(\frac{\Delta}{d}\right)^2}{d\Gamma\left(\frac{2 \Delta}{d}\right)} \mathbf{P}_{-1+\frac{\Delta}{d}} \left(-1+2 (r b)^{d} \right)\,.
\end{equation}
Expanding \eqref{E:solO0} near the asymptotically AdS boundary and using \eqref{E:expansion}  one can evaluate 
\begin{equation}
	\langle O_{\Delta} \rangle = 
	\frac{2 \Gamma\left(\frac{\Delta}{d}\right)^2 \Gamma\left(1-\frac{2 \Delta}{d}\right)}{b^{\Delta} \Gamma\left(1-\frac{\Delta}{d}\right)^2 \Gamma\left(-1+\frac{2 \Delta}{d}\right)}
	(\Lambda b)^{d-\Delta}.
\end{equation}
Since the background is that of a conformal theory then to leading order in $\Lambda/T$ the energy density $\epsilon$ should scale as $\epsilon \sim T^{d+1}$. Using $b=d/(4\pi T)$ and \eqref{E:tracesound}, one obtains \eqref{E:csresult}.

Now I turn to the slightly more difficult problem of computing the bulk viscosity. As stated earlier, instead of the traditional approach to computing the bulk viscosity via the Kubo formula as was done in, for example, \cite{Parnachev:2005hh,Benincasa:2005iv,Benincasa:2006ei,Gubser:2008sz,Gubser:2008yx,Cherman:2009kf},  I will use \eqref{E:zetaidentity} in a background with non vanishing $\partial_{\alpha}u^{\alpha}$. Such black hole backgrounds were constructed in \cite{Bhattacharyya:2008jc,VanRaamsdonk:2008fp,Haack:2008cp,Bhattacharyya:2008mz}: omitting corrections of order $\mathcal{O}(\partial^2)$, they are given by the line element
\begin{subequations}
\label{E:Odsolution}
\begin{equation}
\label{E:MetricO1}
	ds^2 = ds_{(0)}^2+ds_{(1)}^2
\end{equation}
where 
\begin{equation}
	ds_{(0)}^2 = - 2 u_{\mu} dx^{\mu} dr - r^2 f(b r) u_{\mu}u_{\nu} dx^{\mu}dx^{\nu} + r^2 P_{\mu\nu}dx^{\mu} dx^{\nu}
\end{equation}
with 
\begin{equation}
	u_{\mu} = \left(\frac{1}{\sqrt{1-\beta^2}},\frac{\vec{\beta}}{\sqrt{1-\beta^2}} \right)\,,
\end{equation}	
and
\begin{equation}
	ds_{(1)}^2 =\left( 2 r^2 \sigma_{\mu\nu} b F(b r) + \frac{2}{3} r \partial_{a} u^{a} u_{\mu}u_{\nu}
		- r u^{\alpha}\partial_{\alpha} u^{\gamma} \left(P_{\gamma \mu} u_{\nu}+P_{\gamma \nu} u_{\mu} \right)
		\right) dx^\mu dx^\nu
\end{equation}
with
\begin{equation}
\label{E:Fbr}
	F(b r) = -\ln(b r)+\sum_{n=1}^{d-1} \frac{\ln (b r - x_n) x_n^{d-2}}{\sum_{k=0}^{d-2}(k+1)x_n^k}.
\end{equation}
\end{subequations}
In \eqref{E:Fbr}, the $x_n$ are the $d-1$ roots of the polynomial $\sum_{n=0}^{d-1} x^n$,
and $b$ and $u^\mu$ must satisfy
\begin{equation}
\label{E:conservation}
	\partial_{\mu} \left( b^{-d} \left(d u_{\mu}u^{\nu}+\delta_{\mu}^{\phantom{\mu}\nu}\right) - b^{-(d-1)} \sigma_{\mu}^{\phantom{\mu}\nu} \right) = \mathcal{O}(\partial^2).
\end{equation}
Note that $b$ and $u^{\mu}$ now depend on the transverse coordinates $x^\mu$. Put differently, \eqref{E:Odsolution} is a solution to \eqref{E:Einstein} to order $\mathcal{O}(\partial^2)$ (as long as the boundary theory stress tensor is conserved, as in \eqref{E:conservation}) and describes a configuration of the boundary theory with a non trivial flow. See \cite{Rangamani:2009xk} for a review.

With the $\mathcal{O}(\partial)$ solution to \eqref{E:Einstein} at hand one is left with solving \eqref{E:scalareom} in the background \eqref{E:Odsolution}. Similar to the notation in \eqref{E:MetricO1} this solution can be written in the form
\begin{equation}
\label{E:phitotal}
	\phi = \phi^{(0)} + \phi^{(1)}\,,
\end{equation}
where $\phi^{(0)}$ solves  the scalar equation up to $\mathcal{O}(\partial)$. Since at this order the metric is nothing but a boosted version of \eqref{E:AdSSS}, $\phi^{(0)}$ is given by equation \eqref{E:solO0},
\begin{equation}
\label{E:scalarO0}
	\phi^{(0)}(r b) = (\Lambda b)^{d-\Delta} \frac{2\Gamma\left(\frac{\Delta}{d}\right)^2}{d\Gamma\left(\frac{2 \Delta}{d}\right)} \mathbf{P}_{-1+\frac{\Delta}{d}} \left(-1+2 (r b)^{d} \right)\,.
\end{equation}
Inserting \eqref{E:phitotal} and \eqref{E:scalarO0} into \eqref{E:scalareom} I find that $\phi^{(1)}$ must satisfy
\begin{multline}
\label{E:scalarO1}
	\frac{ \partial_r \left( r (1-(r b)^{d}) \partial_r \phi^{(1)} \right)}{r^{d-1} b^{d-2}} +\Delta(\Delta-d)\phi^{(1)} = 
	 \\
	 \frac{4 \Gamma\left(\frac{\Delta}{d}\right)^2}{d\Gamma\left(\frac{2 \Delta}{d}\right)}
	r^{-(d-1)/2} \partial_r \left(r^{(d-1)/2} \mathbf{P}_{-1+\frac{\Delta}{d}} \left(-1+2 (r b)^{d} \right)\right)
	 u^{\alpha}\partial_{\alpha} 
	\left( (\Lambda b)^{d-\Delta} \right)
\end{multline}
where \eqref{E:conservation} was used to trade the gradient of the velocity field with the directional derivative of the energy density. After some guesswork, one finds that
\begin{equation}
\label{E:phi1sol}
	\phi^{(1)}(rb) = R(r b) \mathbf{P}_{-1+\frac{\Delta}{d}} \left(-1+2 (r b)^{d} \right) + C \mathbf{Q}_{-1+\frac{\Delta}{d}} \left(-1+2 (r b)^{d+1} \right)
\end{equation}
with
\begin{equation}
	R(x) = \frac{2 \Gamma\left(\frac{\Delta}{d}\right)^2}{d\, \Gamma\left(\frac{2 \Delta}{d}\right)}
	b u^{\alpha}\partial_{\alpha} 
	\left( (\Lambda b)^{d-\Delta} \right)
 	 \int_x^{\infty} \frac{y^{d-2}}{y^{d}-1}dy
\end{equation}
and $C$ a constant is a solution to \eqref{E:scalarO1} which vanishes at the asymptotically AdS boundary. The function $R(x)$ can be written explicitly in terms of logarithms. Requiring $\phi^{(1)}$ to be finite at $r=1/b$ implies
\begin{equation}
	C=-\frac{4}{d^2}\frac{\Gamma\left(\frac{\Delta}{d}\right)^2}{\Gamma\left(\frac{2 \Delta}{d}\right)}
	b u^{\alpha}\partial_{\alpha} 
	\left( (\Lambda b)^{d-\Delta} \right).
\end{equation}
Expanding \eqref{E:phitotal} as in \eqref{E:expansion} one can extract the gradient corrections to $\langle O_\Delta \rangle$. Inserting this expression into \eqref{E:zetaidentity} and using \eqref{E:conservation} one obtains \eqref{E:zetaresult}.

One would have hoped that using this method the second order transport coefficients of the nonconformal theory (see, for example, \cite{Romatschke:2009kr}) could be computed analytically. Unfortunately, it is difficult to solve the equations of motion for the $\mathcal{O}(\partial^2)$ corrections to $\phi$. A similar problem arises if one tries to compute the bulk viscosity at finite chemical potential from the solutions in \cite{Erdmenger:2008rm,Banerjee:2008th,Torabian:2009qk}.

\section*{Acknowledgements}
I would like to thank S. Gubser, C. Herzog and I. Klebanov for useful discussions and A. Nellore for discussions and initial collaboration on this project. This work was supported in part by the Department of Energy under Grant No. DE-FG02-91ER40671 and by the NSF under award number PHY-0652782.

\bibliographystyle{JHEP}
\bibliography{Zeta}

\end{document}